\documentclass{rspublic}
\usepackage{epsfig,graphics,amssymb,amsmath,verbatim,color}

\begin{document}

\title[The transient swimming of a waving  sheet]{The transient swimming of a waving  sheet}

\author[O. S. Pak and E. Lauga]{On Shun Pak and Eric Lauga\footnote{Corresponding author (elauga@ucsd.edu)}}

\affiliation{
Department of Mechanical and Aerospace Engineering, 
University of California San Diego,
9500 Gilman Drive, La Jolla CA 92093-0411, USA.}

\label{firstpage}

\maketitle

\begin{abstract}{swimming problem; low Reynolds number; locomotion; micro-organisms; cell motility}
Small-scale  locomotion plays an important role in biology. Different modelling approaches have been proposed in the past. The simplest model is an   infinite inextensible two-dimensional waving sheet, {originally introduced by Taylor}, which serves as an idealized geometrical model for both spermatozoa locomotion and ciliary transport in Stokes flow. 
Here we complement classic steady-state calculations by deriving the transient low-Reynolds number  swimming speed of such a waving sheet when starting from rest (small-amplitude initial-value problem). We also determine the transient fluid flow in the `pumping' setup where the sheet is not free to move but instead generates a net fluid flow  around it. The time scales for these two problems, which in general govern transient effects in transport and locomotion at low Reynolds numbers, are also derived using physical arguments.
\end{abstract}

\section{Introduction}

The locomotion of microorganisms plays a vital role in biology.   Examples include the locomotion of mammalian spermatozoa during reproduction, or the swimming of  bacteria and algae to locate  better nutrient sources (Childress 1981; Bray 2000).  Microorganisms adopt different swimming strategies from those exploited by larger animals. For  humans, fish, or birds, swimming and flying are usually accomplished by imparting momentum to the fluid. This  strategy, however, no longer works in the world of microorganisms, where inertia plays little role and viscous damping is paramount. The Reynolds number, a dimensionless parameter characterizing the ratio of  inertial to viscous forces  in the surrounding fluid,  ranges from $10^{-6}$ for flagellated bacteria to $10^{-2}$ for spermatozoa (Brennen \& Winet 1977). Many of such small organisms propel themselves by propagating progressive waves along their flagella. The geometrical characteristics of these microorganisms and the waves they propagate have been reviewed by Brennen \& Winet (1977).  For example, for animal spermatozoa, typical wavelengths range from $11-65 \3 \mu$m and wave amplitudes range from $4-15 \3  \mu$m, while the  ratio  between their  swimming speeds and the wave speeds range from  0.07 to 0.3.
Taylor (1951) initiated the studies on hydrodynamics of microorganisms by modelling flagella swimming. After his pioneering work, much progress has been made towards addressing the basic mechanisms of  propulsion for swimming microorganisms, and we refer to  Brennen \& Winet (1977) and Childress (1981) for  detailed reviews.  In this paper, we focus on unsteady effects in low-Reynolds number locomotion.  
We start below with a brief overview of relevant works before summarizing the approach and outline of  our paper.

In 1951, Taylor first showed that self-propulsion of a waving sheet is possible in a viscous fluid in the absence of inertia. The sheet moves in a direction opposite to that of the wave propagation, and the steady state swimming speed was computed explicitly to be $a^2k^2c/2$ {at leading order in the wave amplitude}, where $a$, $k$ and $c$ are the wave amplitude, wave number and wave propagation speed respectively.  Drummond (1966) extended Taylor's results to larger amplitudes of motion. Reynolds (1965) and Tuck (1968) included inertial effects in the analysis and found that the swimming speed decreases with Reynolds number. In the inertial  realm, Wu (1961, 1971) also considered the problem of a waving sheet, but with finite chord and in the inviscid limit, as a model for fish propulsion. Recently, Childress (2008) discussed the nature of the high Reynolds number limit of Taylor's swimming problem and applied the results to the idea of recoil swimming, where propulsion is achieved by movements of the center of mass and center of volume of the body.

Despite its simplicity, the waving sheet  serves as an idealized geometrical model for both spermatozoa locomotion and ciliary transport. Cilia are short flagella beating collaboratively to produce fluid motion. They are important in many biological transport processes, for example, the transport of mucus in the respiratory tract of humans.  Blake (1971) represented the envelope formed by the cilia tips as an infinite impenetrable oscillating surface, assuming that the cilia are sufficiently closely packed together.  The use of an infinite sheet to model finite length organisms is particularly well justified for elongated and flat organisms such as {\it Paramecium} or {\it Opalina} (Blake 1971). 
If the waving sheet is not allowed to move, there will be a net flow of the fluid in one direction. In this case, the sheet acts as a pump for transporting the fluid. We shall refer this as the `pumping problem' in this paper. In addition, when a second waving sheet is present above the original waving surface, one recovers the problem of  peristaltic pumping, a useful fluid-transport mechanism in physiology and many industrial processes. Readers are referred to reviews on this subject (Jaffrin \& Shapiro 1971; Fauci \&  Dillon 2006).

Another scenario of interest is the propulsion of microorganisms near solid boundaries. Natural microorganisms often swim in narrow passages such as the swimming of spermatozoa in the cervix (Suarez \& Pacey 2006). In addition, during most laboratory examinations, the presence of coverslips imposes solid boundaries near the microorganisms. 
Reynolds (1965) adopted Taylor's swimming sheet model to study swimming near solid walls for small-amplitude waving motion; another approximation, the long wavelength limit, was considered by Shack 
\& Lardner (1974) and Katz (1974). The oscillating wall models by Smelser \textit{et al.} (1974) and Shukla \textit{et al.} (1988) and the layered fluid medium model by Shukla \textit{et al.} (1978) were also proposed to study the interaction between the cervical wall and the cell.

The fact that many biological fluids are non-Newtonian has also received considerable attention. Chaudhury (1979) first extended Taylor's swimming problem to viscoelastic fluids. The same problem was then considered by Sturges (1981) using a more rigorous integral constitutive equation. Fulford \textit{et al.} (1998) modified the resistive force theory  to model the swimming of a spermatozoon in a general linear viscoelastic fluid. Recently, Lauga (2007) revisited Taylor's original calculation using more realistic, nonlinear non-Newtonian fluid models.

This brief literature review  shows that most studies in small-scale biological locomotion focus on solving for the swimming speed of a model organism.  
Since previous work derived solutions for steady-state swimming only, we propose in this paper to go beyond the steady limit and study a prototypical   time-varying situation, namely the initial-value problem of a   model microorganism (Taylor's waving sheet) starting from rest. Physically, such a process is governed by the  small time scale necessary for vorticity created at the swimmer surface to propagate diffusively into the fluid, and belongs to the general class of unsteady Stokes problems.

{As expected, this transient swimming process is also dependent upon the development of the propagating wave from rest; we present here a general analytical treatment for a waving sheet which develops its frequency (or phase speed) from rest to the steady state in an arbitrary manner. An analytical formula describing the general transient propulsion speed of the sheet is then derived, complementing  Taylor's well-known steady state solution.  We also solve the transient pumping problem, where the sheet is not free to move, but instead entrains the surrounding fluid in an unsteady manner.  Unlike their steady counterparts, these two problems are not equivalent in the unsteady case because of time-dependent inertial forces.

The paper is structured as follows. In \S 2, the swimming sheet problem is mathematically formulated with the appropriate non-dimensionalization, governing equations, and boundary conditions. In \S 3, the calculations at first and second order are presented. The results of \S 3 are then applied to study the transient pumping problem (\S 4) and the swimming problem (\S 5). Finally, a physical discussion of our results and a derivation of the time scales involved in transient low-Reynolds number swimming is offered in \S 6.

\section{Formulation}
We consider here an infinite sheet swimming in an incompressible fluid, similar to the model proposed by Taylor (1951). We also allow the wave of displacement along the sheet to include not only normal but also tangential motion  (Blake 1971; Childress 1981). Here, the waving motion is observed in the frame moving at the unknown swimming speed. As the swimming speed is time-dependent during the transient motion, the reference frame is non-inertial and hence a fictitious (inertial) force has to be introduced (see below). In the moving frame, the position of material points, $(x_m,y_m)$, on the waving sheet is written as
\begin{align}
 x_m(x,t)&=x+a \hat{A} \cos (kx-\omega(t) t-\phi), \notag \\ 
 y_m(x,t)&=b  \hat{A} \sin (kx-\omega(t) t),
 \end{align}
where $a$ and $b$ are dimensionless, $\hat{A}$ is a typical wave amplitude and $\phi$ is the phase difference between the longitudinal and transverse motion. The angular frequency $\omega(t)$ is an arbitrary function of time which initially starts from zero and eventually reaches a steady state value of $\omega_\infty$.  The functions describe a traveling wave with tangential motion of wavelength $\lambda=2 \pi / k$ moving in the positive $x$ direction at a time-varying speed $c = \omega (t) /k$. We keep $\omega(t)$ arbitrary in the  analysis below to derive general formulas describing the transient net flow and transient swimming velocity. Specific examples for $\omega(t)$ will then be considered for illustration. The transverse traveling wave considered by Taylor (1951) is obtained when $a=0$, and the $b=0$ case represents a longitudinal traveling wave. The unknown swimming speed of the sheet is denoted by $U(t)$ and is assumed to occur in the direction opposite to that of the wave propagation (see figure~\ref{fig1} for notation).

\begin{figure}[t]
\centering
\includegraphics[width=0.5\textwidth]{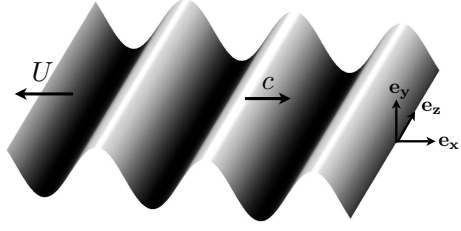}
\caption{Two-dimensional swimming sheet and notation.}
\label{fig1}
\end{figure}

\subsection{Non-dimensionalization}
We non-dimensionalize lengths by $1/k$ and velocities by the wave speed $c$. The time scale in this problem shall characterize changes in velocity in the fluid during the transient motion.  The appropriate, physically-motivated time scale, is that required for the vorticity created by the start-up of the sheet to propagate diffusively into the fluid over the characteristic length scale, and therefore time is non-dimensionalized by $1/k^2 \nu$, where $\nu$ is the kinematic viscosity of the fluid. For a wavelength on the order of 10 $\mu$m, and in water, the time scale considered is on the order of milliseconds. The angular frequency, $\omega(t)$, is non-dimensionalized by its steady state value, $\omega_\infty$. The dimensionless quantities (starred) are summarized as follows, with $\rho$ being the density of the fluid,
\begin{align}
\textbf{x}^* &= \frac{\textbf{x}}{1/k} ,\quad 
t^* = \frac{t}{1/(k^2\nu)} , \quad \textbf{u}^*=\frac{\textbf{u}}{c},\quad 
p^*=\frac{p}{\rho\nu ck}, \quad \omega^*=\frac{\omega}{\omega_\infty},
\end{align}
and the Reynolds number is given by  $Re={c}/{\nu k}={\omega_\infty}/{\nu k^2}$.
The dimensionless position of material points on the sheet is written as
\begin{align}
 {x_m}^* (x^*,t^*)&=x^*+\epsilon a \cos (x^*-Re \ \omega^*(t^*) t^*-\phi), \notag\\
 {y_m}^* (x^*,t^*)&=\epsilon b \sin (x^*-Re \ \omega^*(t^*) t^*),
 \end{align}
 where $\epsilon=\hat{A}k$. We assume that the wave amplitude is much smaller than the wavelength, and derive the results in the limit where $\epsilon \ll 1$.

\subsection{Governing Equation}
We consider an incompressible Newtonian fluid surrounding the sheet. Since we have a two-dimensional setup, the continuity equation, $\nabla \cdot \bold{u}=0$, is satisfied by introducing the stream function $\psi (x,y,t)$ such that $u=\partial \psi / \partial y$ and $v=-\partial \psi / \partial x$, where ${\bf u}=u{\bf e}_x + v {\bf e}_y$. In the laboratory frame, the governing equation is the Navier-Stokes equation
\begin{align}
\rho \frac{\partial \bold {u}}{\partial t} + \rho \left(\bold{u} \cdot \nabla\right) \bold{u}=-\nabla p+\mu \nabla^{2} \bold{u},
\end{align}
where $p$ is the pressure and $\mu$ the dynamic viscosity of the fluid. In the non-inertial frame considered in this paper, as we accelerate with the swimming sheet, a time dependent uniform fictitious force $\bold{F}(t)=(F_x(t), 0, 0)$ has to be introduced into the equation as
\begin{align}
\rho \frac{\partial \bold {u}}{\partial t} + \rho \left(\bold{u} \cdot \nabla \right) \bold{u}=-\nabla p+\mu \nabla^{2} \bold{u}+\bold{F}.
\end{align}
Since the swimming sheet accelerates uniformly in the $x$-direction, the non-zero component of the fictitious force, $F_x$, is uniform (independent of space) and is a function of time only. The non-dimensionalized form of the equation (with non-dimensionalization described above) is given by
\begin{align}
\frac{\partial \bold{u}^*}{\partial t^*} + Re \left(\bold{u}^* \cdot \nabla^* \right) \bold{u}^*=-\nabla^* p^*+\nabla^{*2} \bold{u}^*+\bold{F}^*.
\end{align}
For small-scale biological locomotion, we consider the low Reynolds number limit $Re \rightarrow 0$, where the convective term vanishes, resulting in the unsteady Stokes' equation
\begin{align}
\frac{\partial \bold{u}^*}{\partial t^*}=-\nabla^* p^*+\nabla^{*2} \bold{u}^*+\bold{F}^*.
\end{align}
Upon taking the curl of the equation, both the pressure gradient and the fictitious force terms vanish, resulting in an equation for the  $z$-component of the vorticity, ${\omega}^*$.
With the relation ${\omega}^*=-\nabla^{*2}\psi^*$, the equation for the stream function is given by
\begin{align}
\left(\frac{\partial}{\partial t^*}-\nabla^{*2} \right) \nabla^{*2} \psi^* = 0.
\end{align} 
Hereafter, we shall mostly deal with the dimensionless quantities, and therefore the stars will be omitted for simplicity.

\subsection{Boundary Conditions}
The unknown swimming velocity of the sheet is denoted by $-U (t)  \bold{e_x}$. In the frame moving with the swimming sheet, the velocity of the fluid in the far field ($y=\infty$) is therefore given by $U (t)  \bold{e_x}$. Hence, the far-field boundary conditions are
\begin{align}
\nabla \psi \mid_{(x_m,\infty)}=U(t)  \bold{e_y}.
\end{align}
On the swimming sheet, the boundary conditions are given by the velocity components of a particle of the sheet
\begin{align}
\nabla \psi \mid_{(x_m,y_m)}= \epsilon f(t) \left[ b \cos \left(x-Re \cdot \omega(t) t \right)  \bold{e_x} + a \sin \left(x-Re \cdot \omega(t) t -\phi \right)  \bold{e_y} \right],
\end{align}
where $f(t)$ is a function of time, defined by $f(t)=d\left(\omega(t) t\right)/dt$.
The conditions simplify to
\begin{align}
\nabla \psi \mid_{(x_m,y_m)}= \epsilon f(t) \left[ b \cos \left(x \right)  \bold{e_x} + a \sin \left(x -\phi \right)  \bold{e_y} \right],
\end{align}
in the low Reynolds number limit ($Re \rightarrow 0)$. 
Note that $Re={\omega_\infty}/{\nu k^2}$ can be re-written as $Re=[1/{\nu k^2}]/[1/{\omega_\infty}]$, and can therefore be interpreted as the ratio between the  relevant time scale for viscous diffusion into the fluid and the typical time scale of the wave (its period). The $Re \rightarrow 0$ limit physically means that the diffusion of vorticity occurs much faster than the propagation of the wave, and therefore the wave appears to be stationary {on the time scale where} viscous diffusion is taking place.

\section{Analysis}
In this section, we seek regular perturbation expansions for the stream function, swimming speed and the pressure in powers of $\epsilon$ in the form
\begin{align}
\{\psi, U, p\} &=\epsilon \{\psi_{1}, U_1, p_1\}+\epsilon^2 \{\psi_{2}, U_2, p_2\}+ \ldots.
\end{align} 
The order $\epsilon$ and order $\epsilon^2$ solutions are presented in the following subsections.

\subsection{First-order solution}
At order $\epsilon$, the governing equation is given by
\begin{align}\label{order1}
\left(\frac{\partial}{\partial t}-\nabla^{2}\right) \nabla^{2} \psi_1 = 0.
\end{align}
Expanding the boundary conditions on the sheet using Taylor expansion, they become  at order $\epsilon$ 
\begin{align}
\psi_{1,y}\mid_{(x, 0)} \ &= \ a f(t) \sin(x-\phi),\\
\psi_{1,x}\mid_{(x, 0)} \ &= \ b f(t) \cos x,\\
\psi_{1,y}\mid_{(x, \infty)} \ &= \ U_1(t),\\
\psi_{1,x}\mid_{(x, \infty)} \ &= \  0.
\end{align}
The initial condition is that the vorticity in the fluid ($\bold{\omega}=-\nabla^{2} \psi$) is initially zero, {\it i.e.} $\bold{\omega}_1\mid_{t=0} \ = \ \left(-\nabla^{2} \psi_1\right)\mid_{t=0} \ = \ 0$.

To allow an easy implementation of the initial condition, we solve the problem using the Laplace transform method. A similar technique has been applied to study the transient solution of Stokes' second problem (Erdogan 2000). {The Laplace transform of the stream function $\psi_1$ is defined by the relation
\begin{align}
\tilde{\psi}_1(x,y,s)=\int_{0}^{\infty} \psi_1 (x,y,t) \re^{-s t} dt,
\end{align}
where $s$ is the Laplace variable and tilde variables represent transformed quantities}. Taking the Laplace transform of the governing equation at this order, equation~\eqref{order1} becomes
\begin{align}
\left(s-\nabla^2  \right) \nabla^2 \tilde{\psi}_1&=0.
\end{align}
The boundary conditions in Laplace domain are given by
\begin{align}
\tilde{\psi}_{1,y}\mid_{(x,0)} \ &= \ a \tilde{f}(s) \sin(x-\phi),\\
\tilde{\psi}_{1,x}\mid_{(x,0)} \ &=  \ b \tilde{f}(s) \cos x,\\
\tilde{\psi}_{1,y}\mid_{(x,\infty)} \ &= \ \tilde{U}_1(s),\\
\tilde{\psi}_{1,x}\mid_{(x,\infty)} \ &= \  0.
\end{align}
The solution satisfying all the boundary condition is found to be
\begin{align}\label{order1_final}
\tilde{\psi}_1= & \  \tilde{U}_1\left(y+\frac{\re^{-\sqrt{s}y}}{\sqrt{s}}\right)+\frac{\tilde{f}(s)}{\sqrt{s+1}-1} a \sin \phi (\re^{-\sqrt{s+1}y}-\re^{-y}) \cos x \notag\\
&+\frac{\tilde{f}(s)}{\sqrt{s+1}-1} \left[(b \sqrt{s+1}+a \cos \phi) \re^{-y}-(b+a \cos \phi) \re^{-\sqrt{s+1}y}\right] \sin x.
\end{align}

\subsection{Second-order solution}
We proceed to the  analysis at order $\epsilon^2$ using a similar procedure. The governing equation at this order is
\begin{align}
\left(\frac{\partial}{\partial t}-\nabla^2 \right) \nabla^2 \psi_2 = 0,
\end{align}
with boundary conditions
\begin{align}\label{BC2}
\psi_{2,y}\mid_{(x, 0)} \ &= \  - a \cos (x-\phi)\frac{\partial^2 \psi_1}{\partial x \partial y}\mid_{y=0}-b \sin x \frac{\partial^2 \psi_1}{\partial y^2} \mid_{y=0},\\
\psi_{2,x}\mid_{(x, 0)} \ &= - a \cos (x-\phi)\frac{\partial^2 \psi_1}{\partial x^2}\mid_{y=0}-b \sin x \frac{\partial^2 \psi_1}{\partial x \partial y} \mid_{y=0},\\
\psi_{2,y}\mid_{(x, \infty)} \ &= \ U_2(t),\\
\psi_{2,x}\mid_{(x, \infty)} \ &= \  0,
\end{align}
where we have used Taylor expansion to obtain the boundary conditions on the moving sheet surface. Using Laplace transform, the governing equation for the stream function becomes
\begin{align}
\left(s-\nabla^2 \right) \nabla^2 \tilde{\psi}_2&=0.
\end{align}
The boundary conditions are now
\begin{align}
\tilde{\psi}_{2,y}\mid_{(x,0)} \ &= \  \tilde{f}(s)\left[b^2\sqrt{s+1} \sin^2 x+ab\frac{s}{\sqrt{s+1}-1}\sin x \sin (x-\phi)-a^2 \cos^2 (x-\phi)\right],\\
\tilde{\psi}_{2,x}\mid_{(x,0)} \ &= 0,\\
\tilde{\psi}_{2,y}\mid_{(x,\infty)} \ &= \ \tilde{U}_2(s),\\
\tilde{\psi}_{2,x}\mid_{(x,\infty)} \ &= \  0.
\end{align}
The solution satisfying all the boundary conditions at this order is obtained to be
\begin{align}\label{order2_final}
\tilde{\psi}_2 =& \ \tilde{U}_2 \left(y+\frac{\re^{-\sqrt{s}y}}{\sqrt{s}}\right)-\frac{\tilde{f}(s)}{2\sqrt{s}} \left(b^2\sqrt{s+1}-a^2+ab\frac{s}{\sqrt{s+1}-1}\cos \phi\right) \re^{-\sqrt{s}y}\notag\\
&+\frac{\tilde{f}(s)}{2(\sqrt{s+4}-2)}\left(a^2 \sin 2\phi+ab\frac{s}{\sqrt{s+1}-1} \sin \phi \right)(\re^{-\sqrt{s+4}y}-\re^{-2y})\sin 2x\notag\\
&+\frac{\tilde{f}(s)}{2(\sqrt{s+4}-2)}\left(a^2 \cos 2\phi+b^2 \sqrt{s+1}+ab\frac{s}{\sqrt{s+1}-1}\right)(\re^{-\sqrt{s+4}y}-\re^{-2y})\cos 2x .
\end{align}

\section{Pumping Problem}

The analysis of \S 3 can first be applied to the pumping problem, where the sheet is not allowed to move and its oscillatory motion entrains a net fluid flow along the $x$-direction. Since the sheet does not move, we are in the laboratory frame of reference and the introduction of the fictitious force is unnecessary in this case. The velocity in the far field is zero for any finite time, which is analogous to that in Stokes' problems; therefore, $U_1=U_2=0$, and we then take the $x$-average of the horizontal velocity (denoted by $\langle...\rangle$), leading to
\begin{align}
\langle \tilde{u}_1 \rangle (y,s) &= \frac{\partial \langle \tilde{\psi}_1\rangle}{\partial y}=0,\\
\langle \tilde{u}_2 \rangle (y,s)&= \frac{\partial \langle \tilde{\psi}_2 \rangle }{\partial y}=\frac{\tilde{f}(s)}{2} \left(b^2\sqrt{s+1}-a^2+ab\frac{s}{\sqrt{s+1}-1}\cos \phi\right) \re^{-\sqrt{s}y}\label{final_Laplace_pumping}.
\end{align}
As expected from the $\epsilon \rightarrow - \epsilon$ symmetry,  no net flow occurs at order $\epsilon$. At order $\epsilon^2$, there is a net fluid flow and the result of equation~\eqref{final_Laplace_pumping} describes the (dimensionless) transient velocity of the driven flow in the Laplace domain. For any finite distance $y$, as time goes to infinity, the average horizontal velocity, by the final value theorem, asymptotes to
\begin{align}
\langle \tilde{u}_2 \rangle (y,t \rightarrow \infty)&= \frac{1}{2} (b^2+2ab \cos \phi-a^2)\label{steady_state_pumping},
\end{align}
which is the appropriate steady-state value (Blake 1971; Childress 1981).
The direction of the net fluid flow is governed by the oscillation mode of the swimming sheet. For a  transverse wave ($a=0$), there is a net fluid flow in the positive $x$ direction, which is the direction of the wave propagation. For a  longitudinal wave ($b=0$), the fluid flows in a direction opposite to the wave propagation. If there is a combination between transverse and longitudinal motion ($a\ne0, b\ne0$), the direction of fluid flow is governed by the competition between the amplitudes of the transverse and longitudinal motion and their phase difference, as described by equation~\eqref{steady_state_pumping}. Notably,  there are values of $a$, $b$ and $\phi$ that will yield a zero net fluid flow, where the propulsion caused by the longitudinal motion exactly balances that produced by the transverse motion.

The detail evolution of the transient flow depends on the development of the propagating wave, described by the function $\omega(t)$ (or $f(t)=d\left(\omega(t)t \right)/dt$). Examples will now be given to illustrate the use of equation~\eqref{final_Laplace_pumping} to determine the transient velocity of particular driven flows in the time domain.

\subsection{Example 1: $\omega(t)=2 \arctan\left(t/T \right)/\pi$}
Here, we consider the angular frequency of the wave which develops according to the function 
$\omega(t)=2 \arctan(t/T)/\pi$, 
where $T$ characterizes the time taken for the angular frequency to reach its steady state value, non-dimensionalized by the viscous diffusion time scale. When $T$ is equal to zero, it represents the case where the wave starts impulsively. Computing the function $f(t)=d(\omega(t)t)/dt$ and its Laplace transform $\tilde{f}(s)$,  the transient velocity of the driven flow in Laplace domain follows from equation~\eqref{final_Laplace_pumping} as
\begin{align}
\langle \tilde{u}_2 \rangle (y,s) =& \  \frac{2 \text{Ci}(s T)\left[ \sin (s T) - s T \cos (s T)\right]+\left[ \cos (s T)+ sT \sin(s T)\right] \left[ \pi-2 \text{Si} (s T)\right]}{2 \pi s} \notag\\
& \ \times \left(b^2\sqrt{s+1}-a^2+ab\frac{s}{\sqrt{s+1}-1}\cos \phi\right) \re^{-\sqrt{s}y},
\end{align}
where \text{Ci} and \text{Si} are the cosine and sine integrals, defined respectively by 
\begin{align}
\text{Ci}(x) &= -\int^\infty_x \frac{\cos t}{t} dt, \ \ \ \text{Si}(x) = \int^x_0 \frac{\sin(t)}{t} dt.
\end{align}

The Laplace transform inversion can be easily  implemented numerically (Valk$\acute{\text{o}}$ \& Abate 2004). The evolution of the transient swimming speed in the time domain is computed at the non-dimensional position $y=1$ and shown in figure~\ref{fig2} for different cases: transverse wave (figure~\ref{fig2}a),  longitudinal wave (figure~\ref{fig2}b), and the combined wave where $a=b$ and $\phi=\pi/4$ (figure~\ref{fig2}c). The parameter $T$ is varied from $0.01$ to $100$ in each case, and as expected an increase in $T$ leads to an increase in the time necessary for the flow speed  to reach its steady state value. 

\begin{figure}
\centering
\includegraphics[width=1\textwidth]{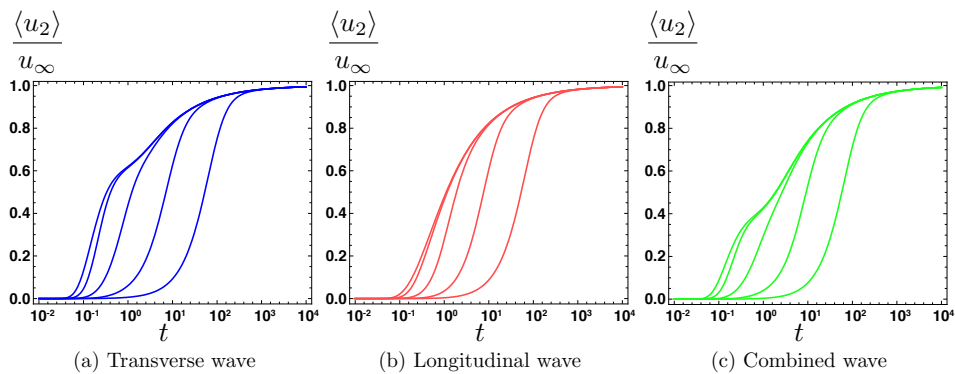}
\caption{Average dimensionless horizontal velocity in the fluid for the pumping problem, $\langle u_2 \rangle / u_{\infty}$, at the non-dimensional position $y=1$, as a function of the dimensionless time, in the case where the wave frequency increases according to 
$\omega (t)=2\arctan(t/T)/\pi$. (Semi-log plot)
(a): transverse wave ($a=0$);
(b): longitudinal wave ($b=0$);
(c): combined transverse and longitudinal wave ($a=b$ and $\phi=\pi/4$).
 In each graph, from left to right: $T=0.01, 0.1, 1, 10, 100$. }
\label{fig2}
\end{figure} 

\subsection{Example 2: Impulsive motion}

We consider now the extreme case where the angular frequency of the propagating wave attains its steady state value instantaneously, {\it i.e.} if $T$ is the dimensionless start-up time, we have $T=0$. The solution to that problem exists as a regular perturbation only in the case where $b=0$, {\it i.e.} when the wave has no transverse amplitude. This is due to the fact that in the  limit $t\to 0$, and as in Stokes' first problem for the impulsive motion of a plate in a viscous fluid, an infinite shear is initially created along the surface of the body. Since a Taylor expansion in the wave amplitude is used to obtain the second-order boundary conditions, equation~\eqref{BC2},  the presence of an infinite shear leads to an infinite boundary condition unless $b=0$, in which case the Taylor expansion does no evaluation  into the fluid domain (see \S 6 for further discussion). 
However, similarly to Stokes' first problem, the solution is  well-behaved  when $b=0$, which is the case we now consider. 
In this case, the function $f(t)$ is a unit step function and its Laplace transform is $\tilde{f}(s)=1/s$. Again, as expected from the $\epsilon \rightarrow - \epsilon$ symmetry, no net fluid flow occurs at order $\epsilon$. Net flow occurs at order $\epsilon^2$. By equation~\eqref{final_Laplace_pumping}, the $x$-average horizontal velocity in Laplace domain in this case is given by
\begin{align}
\langle \tilde{u}_2 \rangle (y,s) &= \frac{\partial \langle \tilde{\psi}_2\rangle}{\partial y}=-\frac{a^2}{2s}\re^{-\sqrt{s}y},
\end{align}
which we inverse Laplace transform, yielding
\begin{align}
\langle u_2 \rangle (y,t) &= -\frac{a^2}{2} {\rm erfc} \left(\frac{y}{2 \sqrt{t}}\right),\label{final_pumping}
\end{align}
where ${\rm erfc}$ is the complementary error function.
\begin{figure}[t]
\centering
\includegraphics[width=0.7\textwidth]{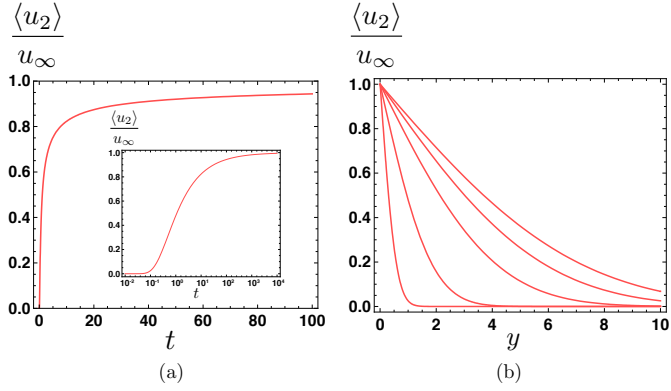}
\caption{(a): Average dimensionless horizontal velocity in the fluid for the pumping problem, $\langle u_2 \rangle / u_{\infty}$, at the non-dimensional position $y=1$, as a function of the dimensionless time for an impulsive longitudinal wave $b=0$. (Inset: same plot on a semi-log scale); (b): Evolution of the average dimensionless horizontal velocity in the fluid in the $y$ direction at different values of dimensionless time for an impulsive longitudinal wave, from left to right: $t=0.1, 1, 5, 10, 15$. }
\label{fig3}
\end{figure} 
The fluid flow occurs in the direction opposite to the wave propagation. The result of equation~\eqref{final_pumping} describes the (dimensionless) transient velocity of the driven flow, and is illustrated in figure~\ref{fig3}. The vorticity perturbation created by the instantaneous motion of the sheet propagates diffusively into the fluid, $y\sim \sqrt{t}$, with a final result very similar to that of Stokes' first problem. For any finite distance $y$, as time goes to infinity, the average horizontal velocity asymptotes to $-a^2/2$. In other words, the entire fluid will eventually be pumped to move with an average horizontal velocity of $-a^2/2$ for very large times. In dimensional form, the average horizontal velocity {is given by}
\begin{align}\label{final1}
\langle u \rangle (y,t)&= -\frac{a^2 \hat{A}^2 k^2 c}{2} {\rm erfc} \left(\frac{y}{2 \sqrt{\nu t}}\right)
\end{align}
at leading order in sheet amplitude.

\section{Swimming Problem}
Next, we consider the swimming problem, where the sheet is free to move. We are now in a frame moving with the sheet, a non-inertial frame of reference (\S2$\,b\)). In this problem, the swimming speed of the sheet is yet to be determined. An additional condition is required to determine $\tilde{U}_1(s)$ and $\tilde{U}_2(s)$ in equations~\eqref{order1_final} and \eqref{order2_final}
respectively. For this purpose, Newton's second law will be applied on the sheet in the $x$-direction. By the periodicity of the problem, we consider the forces (per unit width) acting on one wavelength  of the sheet. The forces are: (i) the horizontal force per unit width acted on the sheet by the fluid (denoted $F_{\rm fluid}$), and (ii) the fictitious force per unit width due to the accelerating reference frame 
(denoted $F_{\rm fictitious}$). In the frame moving with the sheet,  its acceleration is zero, and therefore Newton's second law on the sheet under this non-inertial frame has the form $F_{\rm fluid}+F_{\rm fictitious} = 0$.
The magnitude of the fictitious force, $F_{\rm fictitious}$, is given by the mass per unit width of the sheet times the acceleration of the reference frame. The fictitious force is acting in a direction opposite to the acceleration of the moving frame, so that $F_{\rm fictitious}= (\text{mass of the sheet per unit width}) d U/d t$.
Below, we expand both $F_{\rm fictitious}$ and $F_{\rm fluid}$  in powers of the small parameter $\epsilon$,  and enforce Newton's second law at each order. 

\subsection{First-order solution}
At order $\epsilon$, and in dimensional variables, $F_{\rm fluid}$ is given by
\begin{align}
 \rho \nu c  \int_{0}^{2\pi} (u_{1y}+v_{1x})\mid_{y=0}dx
 =
  \rho \nu c  \int_{0}^{2\pi} \mathcal{L}^{-1}(\tilde{U}_1 \sqrt{s}) dx
 = 2 \pi \rho \nu c  \mathcal{L}^{-1}(\tilde{U}_1 \sqrt{s}),
\end{align}
{where $\mathcal{L}^{-1}$ is the inverse Laplace transform operator.}
The fictitious force, $F_{\rm fictitious}$, at this order reads $
(\rho_{s} h \lambda c k^2 \nu) d U_1/d t,$
where $\rho_{s}$ and $h$ are the density and thickness of the swimming sheet respectively. Applying Newton's second law, we have
\begin{align}
2 \pi \rho \nu c  \mathcal{L}^{-1}(\tilde{U}_1 \sqrt{s}) + \rho_{s} h \lambda c k^2 \nu \frac{d U_1}{d t}=0.
\end{align}
Taking the Laplace transform of the equation yields
\begin{align}
2 \pi \rho \nu c \tilde{U}_1 \sqrt{s} +\rho_{s} h \lambda c k^2 \nu s \tilde{U}_1 =0.
\end{align}
Solving for $\tilde{U}_1(s)$, we have
$
\tilde{U}_1(s)=0,
$
which implies
$
 {U}_1(t)=0.
$
As expected, no self-propulsion occurs at order $\epsilon$. We therefore proceed to the analysis at order $\epsilon^2$.

\subsection{Second-order solution}
A similar analysis is undertaken at this order, with the difference that the order $\epsilon$ pressure, $p_1$, will be required for the calculation of the fluid force, $F_{\rm fluid}$, at  order $\epsilon^2$. The pressure is found by integrating over $x$ the Navier-Stokes equation at order $\epsilon$ in the  horizontal direction. 
{The forces, $F_{\rm fictitious}$ and $F_{\rm fluid}$, are expanded on the sheet using Taylor expansion.} At order $\epsilon^2$, $F_{\rm fluid}$ is given by
\begin{align}
&\rho \nu c  \int_{0}^{2\pi} \left[-(2u_{1,x}-p_1) b \cos x+b \sin x (u_{1,y}+v_{1,x})_y+u_{2,y}+v_{2,x}\right]\mid_{y=0}dx  \notag\\
 &=2\pi \rho \nu c \mathcal{L}^{-1} \left[ \tilde{U}_2\sqrt{s}-\frac{\tilde{f}(s)}{2} \sqrt{s} \left(b^2\sqrt{s+1}-a^2+ab\cos \phi (\sqrt{s+1}-\sqrt{s}+1)\right) \right].
\end{align}
The fictitious force, $F_{\rm fictitious}$, at this order reads
$
(\rho_{s} h \lambda c k^2 \nu) d U_2/d t
$.
Again, applying Newton's second law on the sheet leads to
\begin{align}
2\pi \rho \nu c \mathcal{L}^{-1} \left[ \tilde{U}_2\sqrt{s}-\frac{\tilde{f}(s)}{2} \sqrt{s} \left(b^2\sqrt{s+1}-a^2+ab\cos \phi (\sqrt{s+1}-\sqrt{s}+1)\right) \right] \notag\\
+\rho_{s} h \lambda c k^2 \nu \frac{d U_2}{d t}=0.
\end{align}
Taking the Laplace transform, we have
\begin{align}
\left[ \tilde{U}_2\sqrt{s}-\frac{\tilde{f}(s)}{2} \sqrt{s} \left(b^2\sqrt{s+1}-a^2+ab\cos \phi (\sqrt{s+1}-\sqrt{s}+1)\right) \right]\notag\\
+\frac{\rho_s}{\rho} \frac{h}{(1/k)} s \tilde{U}_2=0.
\end{align}
We define the dimensionless parameter $\mathcal{M}=(\rho_s/\rho)( h k)$, which is the product of the density ratio of the sheet to the fluid and the ratio of the thickness of the sheet to the characteristic length scale (wavelength). The  swimming velocity at order $\epsilon^2$ is then determined, in the Laplace domain, as
\begin{align}\label{swimLaplace}
\tilde{U}_2 (s)= \frac{\tilde{f}(s)}{2(\mathcal{M} \sqrt{s}+1)}\left(b^2\sqrt{s+1}-a^2+ab\cos \phi (\sqrt{s+1}-\sqrt{s}+1)\right) \cdot
\end{align}
The steady state swimming velocity, by the final value theorem, is given by
\begin{align}\label{swimLaplaceSteady}
U_{\infty}=\frac{1}{2}\left(b^2+2ab \cos \phi -a^2\right),
\end{align}
which agrees with  previous steady-state results (Blake 1971; Childress 1981). When $a=0$, it corresponds to Taylor's result of a  transverse traveling wave. The swimming sheet propels itself in a direction opposite to the wave propagation. The $b=0$ limit corresponds to the case of a  longitudinal traveling wave, and the swimming sheet propels in the direction of wave propagation. If there is a combination of transverse and longitudinal motion, similar to the pumping problem, the swimming direction is governed by the competition of the amplitudes of the transverse and longitudinal motion and their phase difference, as described by equation~\eqref{swimLaplaceSteady}. Similarly to the pumping problem, there are values of $a$, $b$ and $\phi$ that will produce a zero propulsion speed.
The transient swimming velocity of the sheet is given by equation~\eqref{swimLaplace} in the Laplace domain for arbitrary function $\tilde{f}(s)$.
Here again, we consider different examples of $\tilde{f}(s)$ and compute the transient swimming velocity in the time domain. 

\subsection{Example 1: $\omega(t)=2 \arctan\left(t/T \right)/\pi$}
Here, we consider the angular frequency of the wave which develops according to the function $\omega(t)=2 \arctan\left(t/T\right)/\pi$, the same function as for  the pumping problem above. Therefore, the functions $f(t)$ and $\tilde{f}(s)$ are the same as  in \S 4a.
By equation~\eqref{swimLaplace}, hence, the swimming velocity in Laplace domain is given by
\begin{align}
\tilde{U}_2 (s) =& \  \frac{2 \text{Ci}(s T)\left[ \sin (s T) - s T \cos (s T)\right]+\left[ \cos (s T)+ sT \sin(s T)\right] \left[ \pi-2 \text{Si} (s T)\right]}{2 \pi s \left(\mathcal{M}\sqrt{s}+1\right)} \notag\\
& \ \times \left(b^2\sqrt{s+1}-a^2+ab\cos \phi (\sqrt{s+1}-\sqrt{s}+1)\right).
\end{align}
The transient swimming velocity in the time domain is obtained by performing a  Laplace transform inversion numerically. 
The results are displayed in figures~\ref{fig4} and~\ref{fig5} for different modes of oscillation: transverse wave (figures~\ref{fig4}a and~\ref{fig5}a), longitudinal wave (figures~\ref{fig4}b and~\ref{fig5}b), and the combined transverse and longitudinal wave where $a=b$ and $\phi=\pi/4$ (figures~\ref{fig4}c and~\ref{fig5}c). 
In figure~\ref{fig4}, the mass ratio is fixed to be $\mathcal{M}=1$ and the wave start-up time $T$ is varied between $0.01$ to $100$. In figure~\ref{fig5}, we fix $T=1$, while the parameter $\mathcal{M}$ is varied between $0.01$ to $100$; the limit where $\mathcal{M}=0$ is also computed and shown as dashed lines in figure~\ref{fig5}. This limit corresponds to the case where the swimming sheet is mass-less, which is the relevant limit for biological organisms.

\begin{figure}[t]
\centering
\includegraphics[width=1\textwidth]{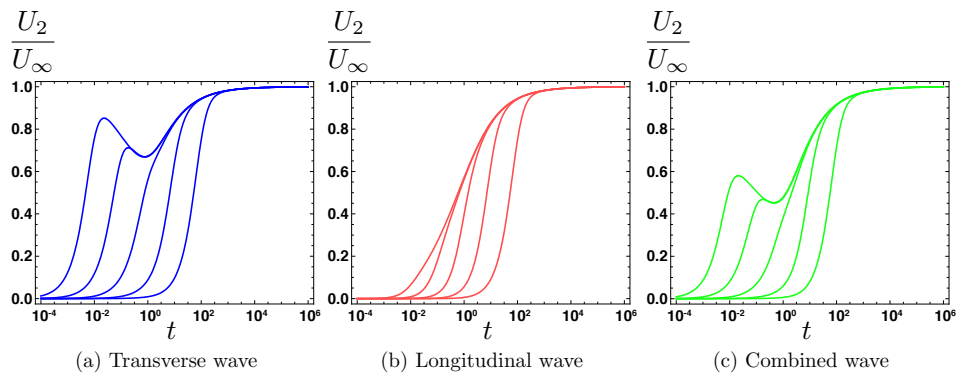}
\caption{Dimensionless swimming speed of the sheet, $U_2/U_{\infty}$, as a function of the dimensionless time, in the case where the dimensionless wave frequency evolves according to  $\omega (t)=2 \arctan(t/T)/\pi$, and for $\mathcal{M}=1$ and different values of the wave start-up time $T$. 
(Semi-log plot). 
(a): transverse wave ($a=0$);
(b): longitudinal wave ($b=0$);
(c): combined transverse and longitudinal wave ($a=b$ and $\phi=\pi/4$).
In each graph, from left to right: $T=0.01, 0.1, 1, 10, 100$. 
}
\label{fig4}
\end{figure}

\begin{figure}[ht]
\centering
\includegraphics[width=1\textwidth]{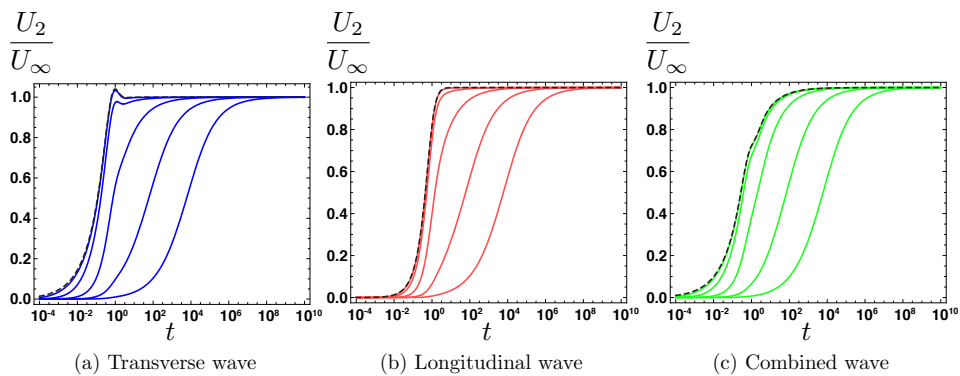}
\caption{
Dimensionless swimming speed of the sheet, $U_2/U_{\infty}$, as a function of the dimensionless time, in the case where the dimensionless wave frequency evolves according to  $\omega (t)=2\arctan(t/T)/\pi$, and for $T=1$ and different values of the mass ratio $\mathcal{M}$. 
(Semi-log plot). 
(a): transverse wave ($a=0$);
(b): longitudinal wave ($b=0$);
(c): combined transverse and longitudinal wave ($a=b$ and $\phi=\pi/4$).
In each graph, from left to right: $\mathcal{M}=0.01, 0.1, 1, 10, 100$. 
The dashed lines (the left-most in each plot) refer to the  case of a mass-less sheet  {\it i.e.} $\mathcal{M}=0$.
}
\label{fig5}
\end{figure}

\subsection{Example 2: $\omega(t)=1-\exp (-t/T)$}
To illustrate the difference between different types of wave start-up, we now consider the case where the angular frequency increases exponentially as: $\omega(t)=1-\exp (-t/T)$. Again, computing the corresponding function $f(t)$ and its Laplace transform $\tilde{f}(s)$,
the swimming velocity in Laplace domain, by equation~\eqref{swimLaplace}, reads
\begin{align}
\tilde{U}_2 (s)= \frac{(1+2sT)}{2s(1+\mathcal{M}\sqrt{s})(1+s T)^2} \left(b^2\sqrt{s+1}-a^2+ab\cos \phi (\sqrt{s+1}-\sqrt{s}+1)\right).
\end{align} 
Here again,  we invert the Laplace transforms numerically. The influence of the dimensionless parameters $T$ and $\mathcal{M}$ on the transient swimming velocity show trends similar to the first example above. The detailed evolution of the swimming velocity is however, and as expected, quantitatively different due to the different time development of the angular frequency. The two different evolution profiles are illustrated  in figure~\ref{fig6}, for the case where $T=1$ and $\mathcal{M}=1$, and for the three different waves.

\begin{figure}
\centering
\includegraphics[width=0.7\textwidth]{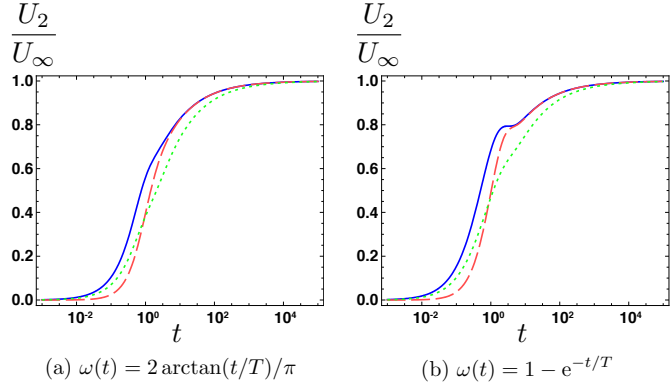}
\caption[leftcaption]{Dimensionless swimming speed of the sheet, $U_2/U_{\infty}$, as a function of the dimensionless time, for $T=1$, $\mathcal{M}=1$, and for two different different wave start-up function $\omega(t)$.(Semi-log plot). 
Blue (solid line): transverse wave ($a=0$);
Red (dashed line): longitudinal wave ($b=0$);
Green (dotted line): combined transverse and longitudinal wave ($a=b$ and $\phi=\pi/4$).
}
\label{fig6}
\end{figure}

In the limit where the swimmer has no mass, $\mathcal{M}=0$, the Laplace transform inversion is simplified and analytical formulas for swimming velocity are obtained for all the three different modes of oscillation. The swimming velocity for a mass-less sheet, for a transverse wave, reads 
\begin{align}
U_2(t)= b^2 \left[\frac{\re^{-t} \sqrt{t}}{2 T \sqrt{\pi }}+\frac{\rm erf \left(\sqrt{t}\right)}{2}-\frac{\re^{-\frac{t}{T}} \left(2 t-3 T-2 t T+2 T^2\right) \rm erf \left(\sqrt{\frac{t(T-1)}{T}}\right)}{4 T^{3/2}\sqrt{T-1}}\right];
\end{align}
whereas for a longitudinal wave, it is
\begin{align}
U_2(t)=-\frac{a^2}{2}\left[1 +\re^{-\frac{t}{T}}\left(\frac{t}{T}-1\right) \right];
\end{align}
for combined motion where, for example,  $a=b$ and $\phi=\pi/4$, the formula is
\begin{align}
U_2(t)=& \ \frac{2+\sqrt{2}}{2} a^2 \left[\frac{\re^{-t} \sqrt{t}}{2 T \sqrt{\pi }}+\frac{\rm erf \left(\sqrt{t}\right)}{2}-\frac{\re^{-\frac{t}{T}} \left(2 t-3 T-2 t T+2 T^2\right) \rm erf \left(\sqrt{\frac{t(T-1)}{T}}\right)}{4 T^{3/2}\sqrt{T-1}}\right] \notag \\
& \ - \frac{\sqrt{2}}{4} a^2 \left[\frac{\sqrt{t}}{T\sqrt{\pi }}-\frac{\re^{-\frac{t}{T}} \left(2 t^{3/2}-4 T \sqrt{t}+T^{3/2}\sqrt{\frac{t}{T}} \right) \rm erfi \left(\sqrt{\frac{t}{T}}\right)}{2 T^{3/2}\sqrt{t}}\right] \notag\\
& \ +\frac{\sqrt{2}-2}{4} a^2 \left[1 +\re^{-\frac{t}{T}}\left(\frac{t}{T}-1\right) \right],
\end{align}
where ${\rm erf}$ is the error function and $\rm erfi$ is the imaginary error function.

\subsection{Example 3: Impulsive motion}
Similarly to the pumping problem, the case of impulsive motion is singular except in the case where $b=0$ for which the impulsive swimming speed is well behaved. In this case,  and as in the pumping problem, the function $f(t)$  is a unit step function and its Laplace transformation is $\tilde{f}(s)=1/s$. Hence, the swimming velocity of the sheet, in Laplace domain, is given by
\begin{align}
\tilde{U}_2(s)=\frac{-a^2}{2s\left(\mathcal{M}\sqrt{s}+1\right)}\label{swimImpulsiveLaplace}\cdot
\end{align}
The inverse Laplace transform of equation~\eqref{swimImpulsiveLaplace} yields the leading order swimming velocity in the time domain
\begin{align}
U_2(t)=-\frac{a^2}{2}\left[{\displaystyle 1-\exp \left(\frac{t}{\mathcal{M}^2} \right) {\rm erfc} \left(\frac{\sqrt{t}}{\mathcal{M}}\right)}\right]\cdot
\end{align}
As in the corresponding pumping problem, the transient swimming velocity involves the complementary error function, ${\rm erfc}(x)$, as illustrated in figure~\ref{fig7}. For large times, the velocity asymptotes to $-a^2/2$. In dimensional form, the  swimming velocity of the sheet is given by
\begin{align}\label{final2}
U(t)=-\frac{a^2}{2}\left[{\displaystyle 1-\exp \left({\frac{\nu \rho^2}{h^2 \rho_{s}^2}} t\right){\rm erfc}\left(\frac{\sqrt{\nu}\rho}{h \rho_s}\sqrt{t}\right)}\right] \hat{A}^2 k^2 c,
\end{align}
at the leading order in the wave amplitude.

\begin{figure}[t]
\centering
\includegraphics[width=0.4\textwidth]{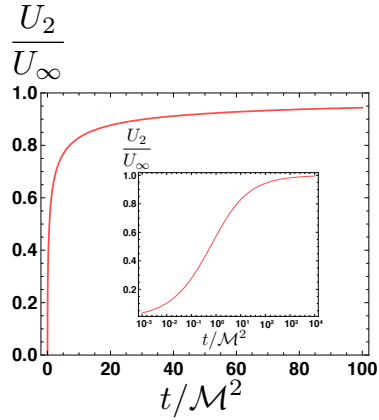}
\caption{Dimensionless swimming speed,  $U_2/U_{\infty}$, for the impulsive longitudinal wave, as a function of the reduced non-dimensionalized time, $t/\mathcal{M}^2$ (Inset: same plot on a semi-log scale).}
\label{fig7}
\end{figure}

\section{Discussion}

In this paper, we have studied two different unsteady Stokes flow problems, and obtained explicit analytical formulas in the Laplace domain for their transient motion. In the pumping problem, the fluid is driven by the waving motion of a fixed sheet, and the Laplace transform of the leading-order dimensionless average horizontal velocity is described by equation~\eqref{final_Laplace_pumping}.
In the swimming  problem, where the sheet is free to move, the leading-order swimming velocity of the sheet is given by equation~\eqref{swimLaplace}. Higher order terms may be obtained in a similar fashion.

We first note the difference between the form of equation~\eqref{final_Laplace_pumping} and equation~\eqref{swimLaplace}. 
In the steady limit, the pumping and swimming problems differ only by a change of reference frame, so the fixed waving sheet pumps a uniform amount of fluid at a velocity equal to minus the steady swimming speed of the free-swimming sheet. In the unsteady case however, the sheet is accelerating, and therefore is subject to additional forces. As a result, the final formula for the swimming speed involves the sheet mass (through the dimensionless parameter $\mathcal{M}$).

To obtain the solution in the time domain, Laplace transform inversion of equation~\eqref{final_Laplace_pumping} and equation~\eqref{swimLaplace} is required.  Analytical formulas have been obtained for several cases where the Laplace transform inversion is simple. It can be further noted that, if a mass-less waving sheet ($\mathcal{M}=0$) is considered in the swimming problem, the Laplace transform inversion is particularly straightforward for the case of longitudinal wave ($b=0$). By equation~\eqref{swimLaplace}, 
we inverse Laplace transform to yield the simple formula
$U(t)=-\epsilon^2 a^2 {f(t)}/{2}$. 

For more complex cases (${\cal M}\neq 0$, $b\neq 0$), the Laplace transform inversion is easily implemented numerically for any admissible $f(t)$ or $\tilde{f}(s)$.  Particular examples have been given for illustration. We  have introduced the dimensionless time $T$  characterizing the time scale over which  the angular frequency  reaches its steady state. As illustrated in figure~\ref{fig2} for the pumping problem, the development of the flow field can be divided into two stages. In the initial stage, the development of the transient velocity depends on the transient motion of the waving sheet, and the relevant time scale is $T$:  the smaller the values of $T$, the shorter the time is required  to reach a given  velocity. For large times, the waving motion of the sheet has effectively reached its steady state, and the subsequent development of the flow field is dominated by the viscous diffusion alone. The values of the parameter $T$ are no longer relevant, and the curves for different values of $T$ collapse onto the same envelop for large times (see figure~\ref{fig2}). In addition, and as could be expected,  the detailed evolution of the net velocity depends on the different waving modes of the sheet. Similar trends are observed in the swimming problem, as displayed in figure~\ref{fig4}. In that case,  since the sheet is free to move, its  mass ---  characterized by the dimensionless parameter $\mathcal{M}$ --- comes into play in its transient swimming velocity, and leads to the presence of a third relevant time scale (see below). As illustrated in figure~\ref{fig5}, the mass of the sheet dictates its acceleration in the initial stage, and the sheet with the smallest mass  accelerates the fastest.

The case where $T=0$, {\it i.e.} where the wave starts impulsively, has also been discussed. We find that, within the framework of the perturbation expansion proposed in this paper, this singular start-up  behavior for the sheet leads to a singular pumping and swimming solution --- except for the case where the wave motion is purely longitudinal, {\it i.e.}  $b=0$.  Physically, as the wave is impulsively starting, it creates an initially infinite shear rate immediately above the sheet. Since the boundary condition for the second order flow is found by the ``sampling'' of the first order flow  by the first-order shape of the sheet (Taylor expansion), any case where the sheet protrudes into the fluid ($b \neq 0$) leads to singular boundary conditions for the second-order flow. 
This singular behavior could be resolved by  incorporating  the advective inertial terms in the Navier-Stokes equation. For example, in the case where $\omega(t)=1-\exp (-t/T)$, it can be shown that, for a  wave with a non-zero transverse amplitude, the convective term scales as $\sim 1/T$ at the sheet position ($y=0$). Hence, for the convective term to be negligible, it is required that $Re/T\ll 1$ or $Re\ll T$, which means that for the results of the present work to be uniformly valid for all time, a transverse wave needs a finite start-up time.
For a longitudinal wave however, the convective term remains order unity for all time, even when the wave is propagated instantaneously, and therefore the impulsive motion problem is well-posed.

A close examination of the results obtained in the impulsive case for the longitudinal wave allows us to understand physically the origin of the time scales involved in the unsteady pumping and swimming processes.
In the pumping problem, the net velocity of the fluid averaged over one wavelength occurs in the direction opposite to the wave propagation, with magnitude given by equation~\eqref{final1}.
The scaling for the time evolution of the fluid velocity in that case is straightforward and similar to that of Stokes' first problem. It is a diffusive scaling, and the fluid below a diffusive front propagating as $y\sim   \sqrt{\nu t}$ into the fluid is pumped roughly at the steady velocity. 
In the swimming problem, the sheet moves in the same direction as the longitudinal wave with a transient propulsion speed  given by equation~\eqref{final2}.
In that case, the scaling for the relevant time scale for start-up of the swimmer arises from a consideration of the balance between the inertial force of the sheet and the shear stresses exerted by the fluid. The inertial force is given by the mass per unit width and per unit wavelength of the sheet times its acceleration. Over one wavelength, the small-amplitude swimmer has a mass per unit width equal to $\rho_s h$, and the acceleration scales as $U_s/t$, where $U_s$ denotes the sheet velocity. On the other hand, the shear stress on the sheet exerted by the fluid is on the order of $\mu U_f/y$, where $U_f$ denotes the fluid velocity and $y\sim \sqrt{\nu t}$ is the typical size in the $y$ direction of  velocity gradients in the fluid. Balancing shear stresses in the fluid with inertia in the sheet  leads to $\rho_s   h U_s/t \sim \mu U_f/\sqrt{\nu t}$. At steady state, we have $U_s\sim U_f$, and the balance suggests a typical time scale, $t_M$, given by  $ t_{M} \sim  \rho_s^2 h^2/\rho^2   \nu $, which is the time scale involved in equation~\eqref{final2}.  Compared to the typical diffusive time scale, $t_D\sim 1/k^2 \nu$, we have $  t_{M}/t_D  = {\cal M}^2$. For typical swimming microorganisms, the ratio of swimmer density to fluid density is about one, and the ratio of thickness $h$ to wavelength $1/k$ is on the order of 0.01 (Brennen \& Winet 1977; Childress 1981), so the steady state swimming problem is reached much earlier than the steady-state pumping problem.

\begin{figure}[t]
\centering
\includegraphics[width=0.4\textwidth]{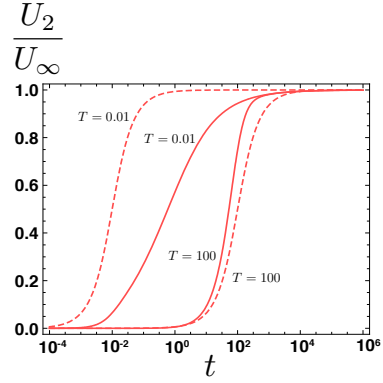}
\caption{Evolution of the dimensionless swimming speed, $U_2/U_\infty$, for the case of longitudinal wave and $\mathcal{M}=1$, given by the result of this paper (solid lines) and the quasi-steady approximation (dashed lines) for different values of $T$ as indicated in the figure.}
\label{fig8}
\end{figure}

Finally, the results of this paper can be compared with those given by a quasi-steady approximation, where the steady-state formulas of Taylor are extended to the unsteady case by replacing the value of the steady-state frequency $\omega_\infty$  by the instantaneous value $\omega(t)$. For illustration, we consider the profile $\omega(t)=2 \arctan\left(t/T \right)/\pi$ in the case  of longitudinal wave, with $\mathcal{M}=1$. The comparison is displayed in  figure~\ref{fig8};  
the solid lines show the evolution of the propulsion speed obtained by our analysis while the dashed lines are obtained using the quasi-steady approximation. Two values of $T$ are employed for illustration. For $T=100$, the development of the wave frequency is slow compared with viscous diffusion, and  the quasi-steady approximation leads to a  reasonable agreement with our results; the inertial effects are relatively unimportant in this limit.
In the other limit for $T=0.01$, the  wave frequency  increases quickly compared to the viscous time scale, and the quasi-steady analysis significantly over-estimates the swimming speed; the inclusion of inertial effects as carried out in this paper is thus critical in this case.

In conclusion, our results  complement Taylor's classical swimming sheet calculation by deriving the transient pumping and swimming motion of the sheet. Generally, the two time scales derived above, which control the startup of the flow surrounding the sheet ($t_D \sim 1/k^2 \nu$), and the startup of the swimmer ($t_M\sim \mathcal{M}^2/k^2\nu$), are expected to govern transient effects in transport and locomotion at low Reynolds numbers.  
In addition, our study could be extended to the case of viscoelastic fluids, for which the relaxation time scale of fluid becomes important, and more complex transient  processes such as the switching of rotation direction in bacterial flagella.

{\begin{acknowledgements}
This work was funded in part by the US National Science Foundation (grants CTS-0624830 and CBET-0746285 to Eric Lauga).
\end{acknowledgements}}

\end{document}